\documentclass[10pt, aps,twocolumn,prc,floatfix,preprintnumbers,superscriptaddress,notitlepage,nofootinbib]{revtex4-1}
\usepackage{dcolumn}
\usepackage{bm}
\usepackage{color}
\usepackage{amssymb}
\usepackage{amsmath}
\usepackage{subcaption}
\usepackage{graphicx}
\usepackage{amsfonts}
\usepackage{slashed}
\usepackage{pstricks}
\usepackage{float}
\usepackage[colorlinks = true,
            linkcolor = blue,
            urlcolor  = blue,
            citecolor = blue,
            anchorcolor = blue]{hyperref}
\usepackage[capitalize]{cleveref}
\usepackage{array}
\allowdisplaybreaks
\usepackage{dcolumn}
\usepackage{epsf}
\usepackage{slashed}
\usepackage[compat=1.1.0]{tikz-feynman}
\usepackage{tikz}
\usetikzlibrary{shapes.geometric}
\usetikzlibrary{calc}
\usepackage{orcidlink}
\usepackage{xspace}
\usepackage{MnSymbol}
\usepackage{multirow}
\usepackage{soul}
\usepackage{slashed}
\usepackage{tikz-feynman}

\definecolor{gray}{rgb}{0.6,0.6,0.6}
\definecolor{darkgreen}{rgb}{0.0, 0.545098, 0.0}
\definecolor{darkblue}{rgb}{0.0, 0.0, 0.545098}
\definecolor{BrickRed}{rgb}{0.8, 0.25, 0.33}
\definecolor{gray}{rgb}{0.6,0.6,0.6}
\definecolor{darkgreen}{rgb}{0.0, 0.545098, 0.0}
\definecolor{mypink1}{rgb}{0.858, 0.188, 0.478}

\def\Fermilab{Theory Division, Fermi National Accelerator Laboratory, P.O. Box 500, Batavia, IL 60510, USA}

\def\Fermilab{Theory Division, Fermilab, P.O. Box 500, Batavia, IL 60510, USA}
\def\IFIC{Instituto de Física Corpuscular (IFIC), CSIC‐Universitat de València, Spain}
\begin{document}

\title{Impact of Two-Body Currents on Semi-Exclusive Lepton–Nucleus Reactions.}

\author{Noemi Rocco}
\email{nrocco@fnal.gov}
\affiliation{\Fermilab}
\affiliation{\IFIC}

\author{Noah Steinberg}
\affiliation{Physics Division, Argonne National Laboratory, Argonne, Illinois 60439, USA}
\affiliation{\Fermilab}

\begin{abstract}
We generalize the spectral-function formalism to describe two-nucleon knockout processes in exclusive kinematics. Significant improvements are introduced both in the treatment of the current operators entering the $\Delta$-current contribution and in the modeling of correlations between the two struck nucleons, including a consistent treatment of isospin dependence and the explicit incorporation of angular correlations. The framework is validated through comparisons with relativistic Fermi-gas calculations and with semi-exclusive electron–nucleus scattering data. Our results demonstrate that an accurate description of nuclear dynamics plays a crucial role in modeling this reaction mechanism. We further present a study of selected electroweak observables relevant to neutrino-scattering experiments. 
\end{abstract}

\maketitle

\section{Introduction}

The contribution to the $\nu A$ cross section from two-nucleon knockout has received considerable attention, as it provides a large fraction of the total cross section at energies around and below $1$ GeV~\cite{Benhar:2015ula,Katori:2013eoa,Barbaro:2016hrt}. In this energy regime, an accurate description of multiple reaction mechanisms is required to correctly reproduce CC0$\pi$ event samples reported by several neutrino oscillation experiments~\cite{Abe:2018CC0pi_Water_T2K,Abe:2020CC0pi_OC_T2K,Abe:2020CC0pi_nu_nubar_T2K,Abe:2020CC0pi_nubar_Water_T2K,Fields:2013MINERvA_QE_like,Betancourt:2017MINERvA_QE_like_Adependence,Abratenko:2020MicroBooNE_CC0piNp_PRD,Abratenko:2020MicroBooNE_CCQE_like_PRL,Abratenko:2023MicroBooNE_multidiff_QElike}. In particular, agreement with data necessitates a consistent treatment of quasi-elastic scattering with single-nucleon emission, multi-nucleon emission, and pion production followed by absorption in the nuclear medium~\cite{Mosel:2018CC0pi,Megias:2019T2K_Oxygen,Kakorin:2021AxialMass,Megias:2019MINERvA,Martini:2022MicroBooNE,FrancoPatino:2024MicroBooNE}. The critical role of these mechanisms in the correct interpretation of neutrino oscillation measurements was first emphasized in connection with the MiniBooNE axial-mass puzzle~\cite{Nieves:2011sc,Nieves:2011pp}.

Ab initio approaches provide a microscopic description of atomic nuclei directly from the underlying interactions among protons and neutrons, without phenomenological adjustments at the many-body level. Within this framework, the Green’s Function Monte Carlo method was the first to enable a fully consistent treatment of one- and two-body currents in electroweak observables, , and that exhibits a sizable effect from the latter, while simultaneously accounting for nuclear correlations. This approach, however, is limited to inclusive observables and fails to fully incorporate relativistic effects~\cite{Lovato:2016gkq,Lovato:2020kba}. To address these limitations, several alternative approaches have been developed that rely on different levels of approximation. Among them, factorized descriptions of the scattering vertex are widely used in the few-GeV regime, including SuSA~\cite{Gonzalez-Rosa:2022ltp,Amaro:2021sec}, relativistic mean-field models~\cite{Gonzalez-Jimenez:2019qhq}, spectral-function approaches~\cite{Benhar:2006wy,Rocco:2015cil}, and short-time approximation schemes~\cite{Andreoli:2024ovl}.
More recently, considerable effort has been devoted to the development of factorized frameworks specifically capable of describing multi-nucleon emission in exclusive kinematics and enabling direct comparisons with experimental data. Electron-scattering measurements have been analyzed within a relativistic global Fermi gas approach in Ref.~\cite{Belocchi:2025eix}. Related developments based on an extension of the Valencia framework construct exclusive final states by applying appropriate cuts to the self-energies of intermediate bosons and include random-phase-approximation effects together with a more realistic parameterization of the nucleon self-energy~\cite{Sobczyk:2024ecl}.

In this work, we present a generalization of the spectral-function formalism required to describe two-nucleon emission processes. To accurately account for the kinematics of the outgoing hadrons, the computational framework has been restructured around a Monte Carlo event generator that samples the full kinematics of both outgoing nucleons and the outgoing lepton on an event-by-event basis. The current operator entering the $\Delta$ contribution have been changed from the  Rarita-Schwinger form to the pure spin-3/2 projector operator and improved using consistent couplings and updated form factors. Furthermore the two-nucleon spectral function distribution used to sample the initial nucleons has been refined by incorporating angular correlations derived from highly realistic Argonne nucleon–nucleon potentials. The resulting framework is designed to be readily interfaced with neutrino event generators, allowing the generated particles to be propagated through intranuclear cascade models.\\
\noindent The paper is organized as follows. In Sec.~\ref{sec:theo:frame}, we present the theoretical framework, detailing the description of two-nucleon correlations and the novel features introduced in the current operators. Section~\ref{sec:res} provides a detailed comparison with semi-exclusive electron-scattering data, highlighting the differences between calculations based on the spectral function and those employing a global Fermi gas model, which retains only statistical correlations. We also present results for neutrino–${}^{12}$C scattering, considering both monochromatic and flux-folded interactions. While no direct comparison with data is performed in this case, we propose experimentally accessible distributions and discuss their sensitivity to nuclear-structure effects.

\section{Theory framework}
\label{sec:theo:frame}
The semi-exclusive differential cross section of two-nucleon emission follows generally from the inclusive definition of the nuclear response tensor, defined as
\begin{equation}
    W^{\mu\nu} = \sum_{f}\langle \Psi_{0}|j^{\mu}|\Psi_{f}\rangle\langle\Psi_{f}|j^{\mu\dagger}|\Psi_{0}\rangle\delta(E_{0} + \omega - E_{f})\, .
\end{equation}
Here, $\Psi_{0}$ and $\Psi_{f}$ are the initial nuclear ground state and an arbitrary final state with energies $E_{0}$ and $E_{f}$ respectively. In the factorization scheme the nuclear current operator $j^{\mu}$ is decomposed as a sum over n-body operators involving an increasing number of nucleons
\begin{equation}
    j^{\mu} = j^{\mu}_{1b} + j^{\mu}_{2b} + \ldots + j^{\mu}_{Ab}\, .
\end{equation}
Isolating the contribution of $j^{\mu}_{2b}$ and inserting a plane-wave ansatz for the two-nucleon final state leads to a factorization of the nuclear matrix element
\begin{equation}\label{eq:factorization}
    \langle\Psi_{f}|J^{\mu}|\Psi_{0}\rangle\rightarrow \sum_{h_{1}h_{2}}[\langle\Psi^{A-2}_{f}|\otimes\langle h_{1}h_{2}|]\Psi_{0}\rangle{}_{a}\langle{p_{1}p_{2}}|j^{\mu}_{2b}|h_{1}h_{2}\rangle\, .
\end{equation}
The first piece of Eq.~\eqref{eq:factorization} is the overlap between the initial A-body state and the A-2 remnant system and is independent of the momentum transfer. Thus it can be computed using non-relativistic nuclear many body theory and becomes an input to the spectral function describing the initial two-nucleon state within the nucleus. 

In inclusive calculations the amplitude in Eq.~\eqref{eq:factorization} would be squared and integrated over the final state nucleon momenta. Semi-exclusive cross sections can be obtained by neglecting this last step, eliminating several simplifications which occur when the final state hadrons go unobserved. The expression for the fully differential two-nucleon knockout cross section reads

\begin{align}
    d\sigma &= V\frac{1}{2E_{k}}\frac{d^{3}k^{\prime}}{(2\pi)^{3}2E_{k^{\prime}}}\frac{d^3p_{1}}{(2\pi)^3 2E_{p_{1}}}\frac{d^3p_{2}}{(2\pi)^3 2E_{p_{2}}}\nonumber\\
    & \times \frac{d^3h_{1}}{2 E_{h_{1}}}\frac{d^3h_{2}}{2 E_{h_{2}}}
    dE\, S(h_1,h_2,E) L_{\mu\nu}W^{\mu\nu}\nonumber\\
    & \times (2\pi)^4\delta^{3}(\mathbf{q} + \mathbf{h}_{1} + \mathbf{h}_{2} - \mathbf{p}_{1} - \mathbf{p}_{2})\nonumber\\
    & \times \delta(\omega - E + 2m_{N}- (E_{p_1}+E_{p_2})).
    \label{eq:diffxsec}
\end{align}
In the above, the incoming lepton has four-momentum
$k^\mu$, while the initial nucleons have four-momenta $h_1^\mu$ and $h_2^\mu$ in an isospin-symmetric nucleus with volume $ V = \frac{A}{\rho} = \frac{2A}{3\pi^2 k_F^3}$. The scattered lepton emerges with four-momentum \( k^{\prime\mu} \), while the two ejected nucleons carry four-momenta \( p_1^\mu \) and \( p_2^\mu \). We assume the nuclear remnant is undetected and integrate over its excitation spectrum $E$. 


In Eq.~\eqref{eq:diffxsec}, $E_{p} =\sqrt{{\bf p}^2 + m^2}$ refers to the on-shell energy of a particle with momentum $|\bf{p}|$. The nuclear spectral function $S(h_1,h_2,E)$ gives the probability of scattering off two nucleons with momentum magnitudes $h_1,h_2$ and leaving the residual nucleus with excitation energy $E$. 
Throughout this work, we compare our results with those obtained within the relativistic Fermi gas (RFG) model, in which the two-nucleon spectral function is defined by retaining only statistical correlations. In this case, it reads
\begin{align}
    S(h_{1},h_{2},E) = \mathcal{N}\Theta(k_F - h_{1})\Theta(k_{F} - h_{2}) \nonumber\\
    \times \delta(E + E_{h_1} + E_{h_2} - 2m_{N} - E_{2p2h})
\end{align}
where $E_{2p2h}$ is a constant binding energy for the two-nucleon pair. The constant $\mathcal{N}$ ensures that the spectral function defined in this way is normalized to 1.

In Eq.~\eqref{eq:diffxsec} the leptonic tensor is defined as
\begin{equation}
    L^{\mu\nu} = g_{\ell}\overline{\sum}_{s,s^{\prime}}J^{\mu\dagger}_{\ell}J^{\nu}_{\ell}\, ,
\end{equation}
where, $g_{\ell}$ is the coupling of the electroweak boson to the lepton, and we average over the initial lepton spins and sum over the final ones. The hadronic tensor $W^{\mu\nu}$ is here defined as
\begin{align}
    W^{\mu\nu} &=\frac{1}{4} g_{h}\sum_{t_{i},s_{i},t_{f},s_{f}}J^{\mu\dagger}_{2b}J^{\nu}_{2b}\, ,
    \label{eq:had_tensor}
\end{align}
where $g_{h}$ is the coupling of the electroweak boson to the hadrons and the factor $1/4$ accounts for the fact that we sum over indistinguishable two-particle two-hole final states. The two-body current is the matrix element of the MEC current operator between the two nucleons in the initial and final states
\begin{equation}
    J^{\mu}_{2b} = \langle p_{1}p_{2}|j^{\mu}_{2b}|h_1 h_2\rangle - \langle p_{2}p_{1}|j^{\mu}_{2b}|h_1 h_2\rangle
\end{equation}
where we have anti-symmetrized the matrix element. The product of the two matrix elements appearing in Eq.~\eqref{eq:had_tensor} leads to four terms. In the inclusive case, thanks to symmetry properties, it is possible to group these terms into two different contributions called direct and exchange. This reduces the number of matrix elements one must compute from four to two. The inclusive cross section further simplifies in the so called $\it{q \,coordinate\,system}$, the one in which the $z$-axis is chosen in the direction of the three-momentum transfer, $\vec{q}$, and the leptonic scattering plane is chosen to lie in the $x-z$ plane~\cite{Rocco:2018mwt}. In this frame the inclusive cross section depends on five response functions $R_{i}$ defined as
\begin{align}
    &R^{cc}=W^{00}\nonumber \\
    &R^{CL}=-\frac{1}{2}(W^{0z} + W^{z0})\nonumber \\ 
    &R^{LL}=W^{zz}\nonumber \\ 
    &R^{T} = (W^{xx} + W^{yy})\nonumber \\ 
    &R^{T^{\prime}} = -\frac{i}{2}(W^{xy} - W^{yx})\, .
    \label{eq:responses}
\end{align}
In the electromagnetic case $R^{T^{\prime}} = 0$. 


In this work we forgo this simplification with the aim to incorporate this work into neutrino event generators in a fully exclusive fashion. As such, the basic building block of our calculation will be $J^{\mu}_{2b}$. An advantage of working with the full two-body current matrix element instead of nuclear response functions is that our calculation is manifestly Lorentz invariant --- the matrix element being the contraction between the hadronic and leptonic currents. More complicated cross sections which depend upon the hadronic final state will entail more responses than those in Eq.~\eqref{eq:responses}. Thus the full contraction in Eq.~\eqref{eq:diffxsec} ensures that the results of our calculation include all tensor elements of $W^{\mu\nu}$, valid for arbitrary observables and in any reference frame.

Throughout this paper, we follow the de Forest prescription of using the shifted momentum transfer $\tilde{q}^{\mu{}} = (\tilde{\omega},\vec{q})$ in the calculation of the currents and the energy conserving $\delta$-function, while using $q^{\mu}$ computed from the leptonic system to compute form factors at the gauge boson-nucleon vertices~\cite{DeForest:1983ahx}. Current conservation is restored by rotating the system so that $\vec{q}$ is along $\hat{z}$ and enforcing that $(J^{V}_{2b})^{z} = \frac{\omega}{q}(J^{V}_{2b})^{0}$, where $(J^{V})^{\mu}$ is the vector piece of the current. We then rotate the system back to the lab frame. This procedure, while arbitrary, is widely adopted and ensures that gauge-invariance is restored~\cite{Benhar:2006wy,Isaacson:2025lyx}. 

\subsection{Meson Exchange current operator}

The MEC operators contain both non-resonant contributions with a pion exchange as well as the excitation of the $\Delta$ resonance
and non-resonant background, required by chiral symmetry.
The operators used in this work follow those of Ref.~\cite{Hernandez:2007qq} updating the structure of the $\Delta$-propagator, some of the form factors and using consistent couplings as it will be discussed in the text. Note that, in the original model of Ref.~\cite{Hernandez:2016yfb},  
the weak excitation of the $D_{13}\ N^\ast(1520)$ resonance is included. Following the argument of Ref.~\cite{Sobczyk:2024ecl}, this contribution is assumed to be negligible for the kinematics considered and therefore, has been neglected. 

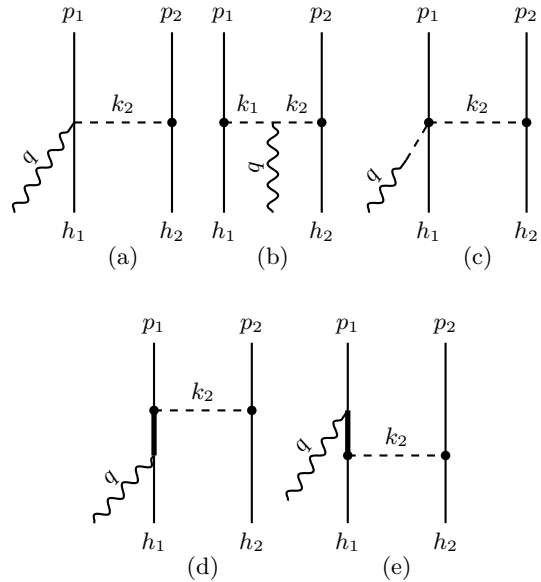
\begin{figure}[ht]
\centering
\begin{tabular}{ccc}

\begin{tikzpicture}[thick,>=stealth]

    \node at (0.65,-1.8) {\small (a)};
  \draw[-] (0,-1.2) -- node[pos=0, below] {$h_{1}$} (0,0.0);=
  \draw[-] (0,0.0) -- node[pos=1, above] {$p_{1}$} (0,1.2);

  \draw[decorate,decoration={snake,amplitude=2pt,segment length=8pt},-]
        (-0.8,-1.2) -- node[above,sloped] {$q$} (0,-0.0);

  \draw[-] (1.3,-1.2) -- node[pos=0, below] {$h_{2}$} (1.3,0.);
  \draw[-] (1.3,0.) -- node[pos=1, above] {$p_{2}$} (1.3,1.2);

  \draw[dashed,-] (0,0.0) -- node[above] {$k_{2}$} (1.3,0.0);

  \foreach \x/\y in { 1.3/0.0} {
    \fill (\x,\y) circle (1.8pt);
  }

\end{tikzpicture}
&
\begin{tikzpicture}[thick,>=stealth]

\node at (0.65,-1.8) {\small (b)};
  \draw[-] (0,-1.2) -- node[pos=0, below] {$h_{1}$} (0,0.0);=
  \draw[-] (0,0.0) -- node[pos=1, above] {$p_{1}$} (0,1.2);

  \draw[decorate,decoration={snake,amplitude=2pt,segment length=8pt},-]
        (0.65,-1.2) -- node[above,sloped] {$q$} (0.65,0);

  \draw[-] (1.3,-1.2) -- node[pos=0, below] {$h_{2}$} (1.3,0.);
  \draw[-] (1.3,0.) -- node[pos=1, above] {$p_{2}$} (1.3,1.2);

  \draw[dashed,-] (0,0.0) -- node[above] {$k_{1}$} (0.65,0.0);
  \draw[dashed,-] (0.65,0.0) -- node[above] {$k_{2}$} (1.3,0.0);

  \foreach \x/\y in { 0/0.0, 1.3/0.0} {
    \fill (\x,\y) circle (1.8pt);
  }

\end{tikzpicture}
&
\begin{tikzpicture}[thick,>=stealth]
\node at (0.65,-1.8) {\small (c)};
  \draw[-] (0,-1.2) -- node[pos=0, below] {$h_{1}$} (0,0.0);=
  \draw[-] (0,0.0) -- node[pos=1, above] {$p_{1}$} (0,1.2);

  \draw[decorate,decoration={snake,amplitude=2pt,segment length=8pt},-]
        (-0.8,-1.2) -- node[above,sloped] {$q$} (-0.3,-0.5);

  \draw[dashed,-]
        (-0.3,-0.5) -- node[above,sloped] {} (0,-0.0);

 \draw[-] (1.3,-1.2) -- node[pos=0, below] {$h_{2}$} (1.3,0.);
  \draw[-] (1.3,0.) -- node[pos=1, above] {$p_{2}$} (1.3,1.2);

  \draw[dashed,-] (0,0.0) -- node[above] {$k_{2}$} (1.3,0.0);

  \foreach \x/\y in { 0/0.0, 1.3/0.0} {
    \fill (\x,\y) circle (1.8pt);
  }
\end{tikzpicture}
\\[1.0em]

\multicolumn{3}{c}{%
  \begin{tabular}{cc}
    \begin{tikzpicture}[thick,>=stealth]
    \node at (0.65,-1.8) {\small (d)};

  \draw[-] (0,-1.2) -- node[pos=0, below] {$h_{1}$} (0,-0.3);
  \draw[-, line width=2pt] (0,-0.3) -- node[right] {} (0,0.3);
  \draw[-] (0,0.3) -- node[pos=1, above] {$p_{1}$} (0,1.2);

  \draw[decorate,decoration={snake,amplitude=2pt,segment length=8pt},-]
        (-0.8,-1.2) -- node[above,sloped] {$q$} (0,-0.3);

  \draw[-] (1.3,-1.2) -- node[pos=0, below] {$h_{2}$} (1.3,0.3);
  \draw[-] (1.3,0.3) -- node[pos=1, above] {$p_{2}$} (1.3,1.2);

  \draw[dashed,-] (0,0.3) -- node[above] {$k_{2}$} (1.3,0.3);

  \foreach \x/\y in { 0/0.3, 1.3/0.3} {
    \fill (\x,\y) circle (1.8pt);
  }

\end{tikzpicture}
    &
    \begin{tikzpicture}[thick,>=stealth]
    \node at (0.65,-1.8) {\small (e)};

  \draw[-] (0,-1.2) -- node[pos=0, below] {$h_{1}$} (0,-0.3);
  \draw[-, line width=2pt] (0,-0.3) -- node[right] {} (0,0.3);
  \draw[-] (0,0.3) -- node[pos=1, above] {$p_{1}$} (0,1.2);

  \draw[decorate,decoration={snake,amplitude=2pt,segment length=8pt},-]
        (-0.8,-0.9) -- node[above,sloped] {$q$} (0,0.3);

  \draw[-] (1.3,-1.2) -- node[pos=0, below] {$h_{2}$} (1.3,-0.3);
  \draw[-] (1.3,-0.3) -- node[pos=1, above] {$p_{2}$} (1.3,1.2);

  \draw[dashed,-] (0,-0.3) -- node[above] {$k_{2}$} (1.3,-0.3);

  \foreach \x/\y in {0/-0.3, 1.3/-0.3} {
    \fill (\x,\y) circle (1.8pt);
  }

\end{tikzpicture}
  \end{tabular}
}
\end{tabular}

\caption{Subset of MEC diagrams contributing to $j^{\mu}_{2b}$. Lines represent nucleons (solid), pions (dashed), $\Delta$s (thick solid), gauge bosons (wavy lines).  Diagrams (a), (b), and (c) are respectively the Seagull, Pion-in-flight, and Pion-pole diagrams, while diagrams (d) and (e) are the forwards and backwards $\Delta$ diagrams. Filled circles at vertices represent the insertion of strong form factors which modify the $\pi NN$ and $\pi N \Delta$ vertices.}
\label{fig:mec_diagrams}
\end{figure}

The MEC are written as the sum of four different contributions
\begin{align}
j^\mu_{2b}= j^\mu_{\pi}+j^\mu_{\rm sea}+ j^\mu_{\rm pole}+ j^\mu_{\Delta}\, ,
\end{align}
A subset of these diagrams is is shown in Fig.~\ref{fig:mec_diagrams} not displaying the contributions where particle 1 and 2 are exchanged.
The diagrams which only include nucleon and pion degrees of freedom comprise the seagull, pion-in-flight, and pion-pole contribution. The expressions used for these terms can be found in Ref.~\cite{RuizSimo:2016rtu}.
The electromagnetic counterparts of the EW currents can be obtained by setting the axial pieces of each current to zero and modifying their isospin dependence. 

The operator containing the $\Delta$-excitation is parametrized as 
\begin{align}
j^\mu_{\Delta}&=\frac{3}{2}\frac{f_{\pi NN} f^\ast}{m^2_\pi} \bigg\{ \Pi(k_2)_{(2)}
\Big[ \Big( \frac{2}{3}\tau^{(2)}-\frac{I_V}{3}\Big)_{\pm}\nonumber\\
&\times F_{\pi NN}(k_2) F_{\pi N \Delta} (k_2) (J^\mu_a)_{(1)}
+\Big(\frac{2}{3}\tau^{(2)}+\frac{I_V}{3}\Big)_{\pm}\nonumber\\ 
&\times F_{\pi NN}(k_2) F_{\pi N \Delta} (k_2) (J^\mu_b)_{(1)}\Big]+(1\leftrightarrow 2)\bigg\}\, .
\label{eq:delta_curr}
\end{align} 
Note that, as detailed in the expressions below, for the $\Delta$ contribution we adopt the sign convention used in Refs.~\cite{Franco-Munoz:2022jcl,Franco-Munoz:2023zoa}, which differs by an overall minus sign from that of Ref.~\cite{RuizSimo:2016rtu}.
In Eq.~\eqref{eq:delta_curr} the operator responsible for pion propagation and absorption, $\Pi(k_{i})$, is given by
\begin{equation}
\Pi(k_{i}) = \frac{\gamma^{5}\slashed{k_{i}}}{k_{i}^{2} - m_{\pi}^2}\, ,
\end{equation}
with $k_{i} = p_{i} - h_{i}$.
The isospin dependence of $j^{\mu}_{\Delta}$ is contained in the Pauli matrices $\tau^{(i)}$ acting on the $i^{th}$ nucleon and the raising-lowering operator
\begin{equation}
    I_{V} = (\tau^{(1)}\times \tau^{(2)})\, ,
\end{equation}
with $\pm\rightarrow x \pm iy$. We use $f_{\pi NN}^2/4\pi = 0.08$, and updated the value of $f^\ast$= 2.15 and 
\begin{equation}
F_{\pi N \Delta}(k)=\frac{\Lambda^2_{\pi N\Delta}}{\Lambda^2_{\pi N\Delta}-k^2}\ ,
\end{equation}
is the strong $\pi N \Delta$ form factor and $\Lambda_{\pi N\Delta}=1200$ MeV.
The $N\rightarrow \Delta$ transitions entering the left (d) and right (e) diagrams of Fig.~\ref{fig:mec_diagrams}, corresponding to $J^\mu_a$ and $J^\mu_b$, respectively
are expressed as
\begin{align}\label{eq:jmua}
J^\mu_a&=(J^\mu_a)_V+(J^\mu_a)_A\ ,\nonumber\\
(J^\mu_a)_V&=k_2^\alpha \tilde{G}_{\alpha\beta}(h+q)
\Big[\frac{C_3^V}{M}
\Big(g^{\beta\mu}\slashed{q}-q^\beta\gamma^\mu\Big)\nonumber\\
&+ \frac{C_4^V}{M^2}\Big(g^{\beta\mu}q\cdot p_\Delta -q^\beta p^\mu_\Delta\Big)\nonumber\\
&+\frac{C_5^V}{M^2}\Big(g^{\beta\mu}q\cdot h -q^\beta h^\mu\Big) \Big]\gamma_5\ ,\nonumber\\
(J^\mu_a)_A&=\Big[\frac{C_4^A}{M^2}\Big(g^{\alpha\mu} q\cdot p_\Delta - q^\alpha p_\Delta^\mu\Big)\nonumber\\
&+C_5^A g^{\alpha\mu} +\frac{C^A_6}{M^2}q^\mu q^\alpha\Big]
k_2^\alpha  
\tilde{G}_{\alpha\beta}(h+q) g^{\beta\mu}\, ,
\end{align}
and
\begin{align}\label{eq:jmub}
J^\mu_b&=(J^\mu_b)_V+(J^\mu_b)_A\ ,\nonumber\\
(J^\mu_b)_V&=\gamma_5 \Big[\frac{C_3^V}{M}
\Big(g^{\alpha\mu}\slashed{q}-q^\alpha\gamma^\mu\Big)\nonumber\\
&+ \frac{C_4^V}{M^2}\Big(g^{\alpha\mu}q\cdot p_\Delta -q^\beta p^\mu_\Delta\Big)\nonumber\\
&+\frac{C_5^V}{M^2}\Big(g^{\beta\mu}q\cdot p -q^\beta p^\mu\Big) \Big] \tilde{G}_{\alpha\beta}(p-q)k_2^\beta ,\nonumber\\
(J^\mu_b)_A&=\Big[\frac{C_4^A}{M^2}\Big(g^{\alpha\mu} q\cdot p_\Delta - q^\alpha p_\Delta^\mu\Big)\nonumber\\
&+
C_5^A g^{\alpha\mu} +\frac{C^A_6}{M^2}q^\mu q^\alpha\Big]
k_2^\alpha  
\tilde{G}_{\alpha\beta}(p-q) g^{\beta\mu}\, .
\end{align}
In Eq.~\eqref{eq:jmub} we have used $\tilde{\Gamma}^{\mu\alpha}(p,q) = \gamma^{0}\Gamma^{\alpha\mu}(p,-q)^{\dagger}\gamma^{0}$ to relate the backwards $\pi N\Delta$ operator $\tilde{\Gamma}(p,q)$ to the forwards $\pi N\Delta$ operator used in Eq.~\eqref{eq:jmua}.

By conservation of the vector current (CVC), the pseudoscalar vector form factor vanishes, $C_6^V=0$. The remaining vector form factors $C^V_{3,4,5}$ are constrained by pion electro-production data as in Ref.~\cite{Hernandez:2016yfb}. In contrast, the axial transition form factors are largely unconstrained: weak single-pion production is typically used to extract information on the nucleon-to-resonance axial couplings. Among these, the term proportional to $C_5^A$ is dominant.

Below we summarize the new form factors introduced and their updated value. Following the pion-pole dominance for the pseudoscalar axial form factor $C_6^A$, partial conservation of the axial current (PCAC) to $C_6^A$ to $C_5^A$ as
\begin{equation}
C_6^A(q^2)= C_5^A(q^2)\,\frac{M^2}{m_\pi^2 - q^2}\,,
\end{equation}
where $q^\mu$ is the lepton four-momentum transfer, and $m_\pi$ is the pion mass. In addition, following Adler's model~\cite{Adler:1968tw}, we set
\begin{equation}
C_3^A(q^2)=0,\qquad C_4^A(q^2) = -\frac{1}{4}\,C_5^A(q^2)\,.
\end{equation}
For the latter we use the expression
\begin{equation}
   C_5^A(q^2) = \frac{C_5^A(0)}{(1-q^2/M_{A\Delta}^2)^2}\, ,
\end{equation}
where the value of the parameters adopted comes from the new fit of Ref.~\cite{Hyper-Kamiokande:2018ofw}
\begin{equation}
    C^A_5(0) = 1.18, \ \ \ \ \ M_{A\Delta} = 950\ \rm{MeV}\, .
\end{equation}

Note that, the de Forest prescription of using $q^{\mu}$ in the form factors and $\tilde{q}^{\mu}$ in the currents has interesting implications for the axial piece of the $\Delta$ current involving $C^{A}_{6}$. PCAC requires that the four-momentum used in the form factor and current are the same, which is obviously violated in this formalism. This enhances some divergences in the $\Delta$ current but does not invalidate the calculation. Further investigation of this effect and the entire De Forest prescription is left to future work.

For the $\Delta$ propagator, we follow Ref.~\cite{Hernandez:2016yfb}
changed it from the Rarita-Schwinger form 
to make sure that only
the pure spin-3/2 projector operator is included and consistent couplings
are adopted~\cite{Pascalutsa:2000kd}. The new propagator is denoted with 
\begin{align}
\tilde{G}^{\alpha\beta}(p_\Delta)=\frac{\tilde{P}^{\alpha\beta}(p_\Delta)}{p^2_\Delta-M_\Delta^2 + i \Gamma(p_\Delta)M_\Delta}\, .
\end{align}

Removing the spurious spin $1/2$ component requires the introduction of 
\begin{align}\label{eq:delta_prop}
    \tilde{P}_{\mu\nu}=& \frac{p^2_\Delta}{M_\Delta^2}P_{\mu\nu}^{\frac{3}{2}} \nonumber \\
    P_{\mu\nu}^{\frac{3}{2}} =&- (\slashed{p}_\Delta +M_\Delta)\Big[g_{\mu\nu}-\frac{1}{3}\gamma_\mu\gamma_\nu\nonumber\\
    &-\frac{1}{3 p_\Delta^2}(\slashed{p}_\Delta \gamma_\mu p_{\Delta\, \nu} + p_{\Delta\, \mu} \gamma_\nu \slashed{p}_\Delta )\Big]\, .
\end{align}

The full effect of using the spin $3/2$ projector of Eq.~\eqref{eq:delta_prop} in the $\Delta$ propagator on exclusive two-nucleon knockout cross sections remains to be studied. At the inclusive level we find the removal of the spurious spin $1/2$ component leads to a reduction of $30\% \,(10\%)$ in the inclusive cross section $\sigma_{2p2h}(E_{\nu})$ at $E_{\nu}$ = 600 MeV (2000 MeV). Consequently, one expects the effects of different the projectors, $\tilde{P}^{\alpha\beta}$, used in the $\Delta$ propagator to be larger at T2K and MicroBooNE energies than at MINER$\nu$A and NO$\nu$A energies.

The possible decay of the $\Delta$ into a physical $\pi N$ state is accounted for by introducing the energy dependent decay width $\Gamma$. Note that, this is modified to account for in medium effects as discussed in Ref.~\cite{Lovato:2023khk}. In contrast with some approaches, we include both the real and imaginary part of the of $\Delta$ propagator. Disregarding the imaginary part of the propagator leads to a strong reduction of the peak and a shift of the strength towards lower energy transfers. The uncertainty arising from this choice will be discussed in future work. 

\subsection{Spectral Function}\label{sec:SpectralFunc}
\begin{figure}
    \centering
    \includegraphics[width=\linewidth]{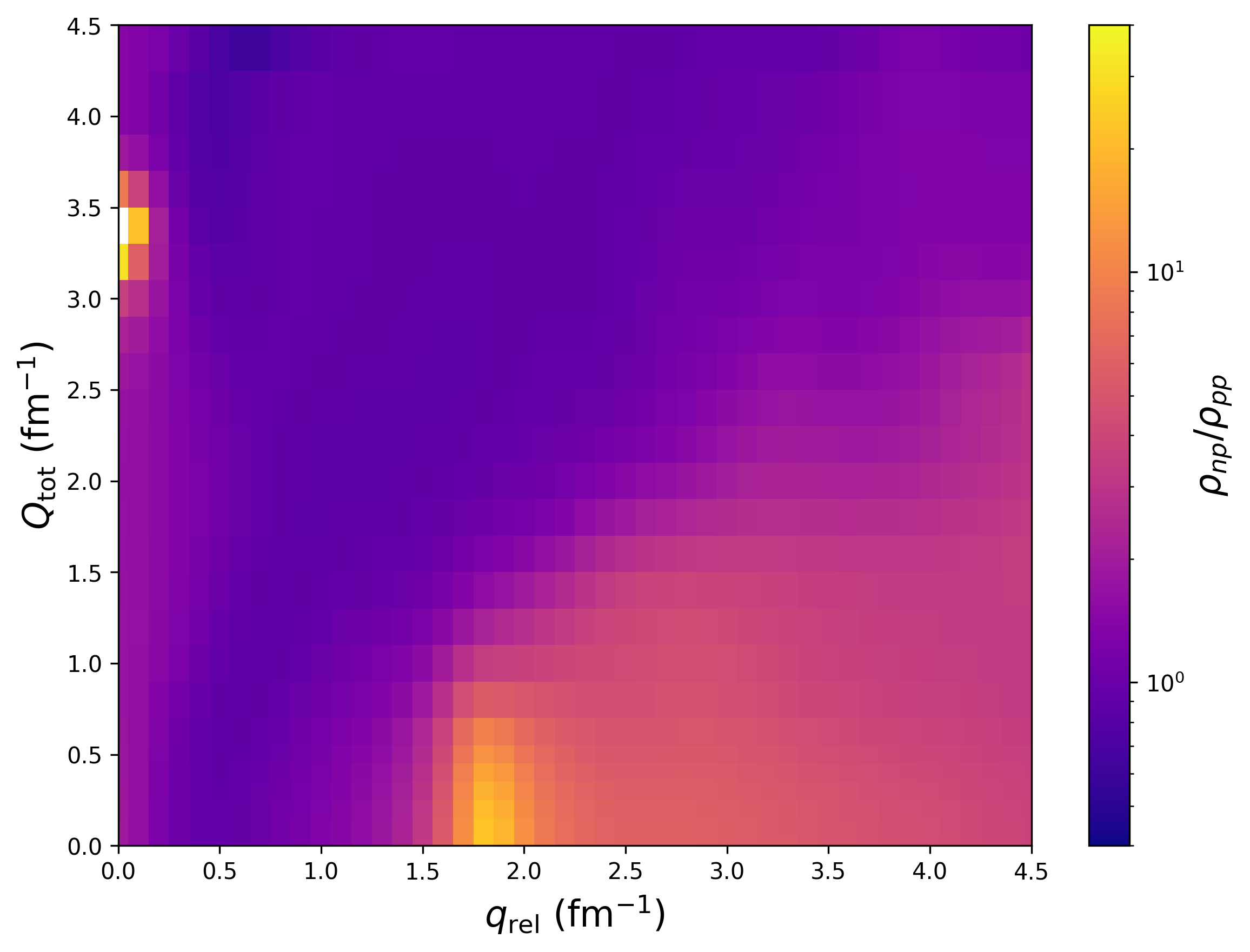}
    \caption{Ratio of normalized two-body momentum distributions for $np$ and $pp$ pairs as a function of $q$ and $Q$. Both distributions have been normalized to 1.}
    \label{fig:qQSF}
\end{figure}

Contributions arising either from squared amplitudes of one-body currents or from the interference between one- and two-body current operators can be expressed in terms of the one-nucleon spectral function, which encodes the properties of the nuclear ground state and of the spectator system.
In the present work, we employ a generalization of this formalism based on the two-nucleon spectral function. Unlike previous studies, we explicitly incorporate, for the first time, the angular dependence of the correlations between the two struck nucleons. In our earlier works, in close analogy with the one-body case, the two-body spectral function was written as the product of a momentum distribution and an energy-conserving delta function. In this approximation, the residual $(A-2)$ system is assumed to be left in a bound state with a fixed binding energy $\bar{B}_{A-2}$, leading to
\begin{align}
S_{S,T}(h_1,h_2,E)
&=
\rho_{S,T}(h_1,h_2)\,
\delta\!\left(
E - B_0 + \bar{B}_{A-2}
- \frac{{\bf Q}^2}{2m_{A-2}}
\right),
\label{eq:SF_approx}
\end{align}
where ${\bf Q}={\bf h}_1+{\bf h}_2$ is the center-of-mass momentum of the struck pair, $B_0$ is the binding energy of the nuclear ground state, and $m_{A-2}$ denotes the mass of the spectator system. The subscript $(S,T)$ labels the spin--isospin channel of the nucleon pair, and $\rho_{S,T}$ denotes the corresponding two-nucleon momentum distribution, which in this approximation depends only on the magnitudes of ${\bf h}_1$ and ${\bf h}_2$.

A key improvement of the present work consists in relating the spectral function to the probability of finding two nucleons inside the nucleus with a given relative momentum
\begin{equation}
{\bf q} = \frac{{\bf h}_1 - {\bf h}_2}{2}
\end{equation}
and center-of-mass momentum
\begin{equation}
{\bf Q} = {\bf h}_1 + {\bf h}_2,
\end{equation}
thereby retaining the full angular dependence of the two-body correlations. The corresponding two-nucleon momentum distribution is defined as
\begin{align}
\rho_{ST}({\bf q},{\bf Q})
&=
\int
d{\bf r}_1^\prime\, d{\bf r}_1\,
d{\bf r}_2^\prime\, d{\bf r}_2\,
d{\bf r}_3 \cdots d{\bf r}_A \;
\nonumber\\
&\times
\Psi_0^\dagger({\bf r}_1^\prime,{\bf r}_2^\prime,{\bf r}_3,\ldots,{\bf r}_A)\,
e^{-i{\bf q}\cdot({\bf r}_{12}-{\bf r}_{12}^\prime)}\,
e^{-i{\bf Q}\cdot({\bf R}_{12}-{\bf R}_{12}^\prime)}
\nonumber\\
&\times
P_{ST}(12)\,
\Psi_0({\bf r}_1,{\bf r}_2,{\bf r}_3,\ldots,{\bf r}_A),
\label{eq:rho_ST}
\end{align}
where ${\bf r}_{12}={\bf r}_1-{\bf r}_2$ and ${\bf R}_{12}=({\bf r}_1+{\bf r}_2)/2$ are the relative and center-of-mass coordinates of the pair, respectively. The operator $P_{ST}(12)$ projects the nucleon pair $(12)$ onto definite spin $S=0,1$ and isospin $T=0,1$. Alternative choices of projectors allow one to isolate contributions from different pair channels.

The normalization
\begin{align}
N_{ST}
=
\int \frac{d{\bf q}}{(2\pi)^3}\,
\frac{d{\bf Q}}{(2\pi)^3}\;
\rho_{ST}({\bf q},{\bf Q})
\label{eq:N_ST}
\end{align}
gives the total number of nucleon pairs in the spin--isospin channel $(S,T)$.

The VMC results used here are obtained with the AV18 two-nucleon interaction and the UIX three-nucleon force~\cite{Wiringa:2013ala}.  
In the \(S=1,\,T=0\) (\(np\)) channel, the tensor operator
\(S_{12}\,\boldsymbol{\tau}_1\!\cdot\!\boldsymbol{\tau}_2\)  
redistributes probability from low relative momentum to 
\(q\simeq 2~\mathrm{fm}^{-1}\), generating the characteristic deuteron-like short-range–correlation (SRC) ridge. This can be seen in Fig.~\ref{fig:qQSF} where we have plotted the ratio $\rho_{np}/\rho_{pp}$ where each momentum distribution has been normalized to one.
At larger momenta, the tail is dominated by central short-range repulsion and spin–isospin correlations, which contribute in all \((S,T)\) channels but do not produce the same strong \(np\) dominance.
Therefore, to construct the two-body momentum distribution with the angular dependence between \(\mathbf{q}\) and \(\mathbf{Q}\) integrated out, we define the two-body spectral function for a given isospin channel \(T\) as
\begin{align}
&S_{ST}(h_1,h_2,\cos\theta_{h_1h_2},E)
= \rho_{ST}({\bf q},{\bf Q})\;\nonumber\\
&\times \delta\!\left(
E - B_0 + B_{A-2}
- \frac{Q^2}{2m_{A-2}}
\right)\,.
\end{align}

In the equation above, the relative and center-of-mass momenta \(q\) and \(Q\) are understood to be the functions of \(h_1\), \(h_2\), and \(\cos\theta_{h_1h_2}\) obtained from the transformation between single-particle and pair variables.

From \(q\) and \(Q\) one may reconstruct the scalar product of the single-particle momenta:
\begin{equation}
{\bf h}_1\!\cdot\!{\bf h}_2
= \frac{Q^2}{4} - q^2.
\end{equation}
Thus, knowing the magnitudes \(|\mathbf{q}|\) and \(|\mathbf{Q}|\) already determines whether the two nucleons tend to be parallel (\(Q^2/4>q^2\)) or back-to-back (\(Q^2/4<q^2\)).  
The distribution \(\rho_{ST}(q,Q)\) therefore encodes both the sign and the strength of their mutual correlation.

Finally, we emphasize that, unlike in some previous works, the isospin summation is treated consistently by isolating each pair's isospin channel both in the spectral function and in the current operator. One can see from Fig.~\ref{fig:qQSF} that dependence of $\rho(q,Q)$ on the isospin of the pair is quite large at high relative momenta, whereas in the Fermi-Gas case the corresponding ratio would be one uniformly.
This consistent treatment is essential for obtaining realistic predictions in exclusive channels and for identifying the characteristic angular patterns associated with different nucleon-pair types. In the next section, we assess the role of correlation effects on a variety of observables relevant to electron and neutrino scattering experiments by comparing our results with those obtained within a Fermi-gas calculation.

 
\section{Results}\label{sec:res}

\begin{figure}[ht]
     \centering
     \begin{subcaption}
         \centering
         \includegraphics[width=\linewidth]{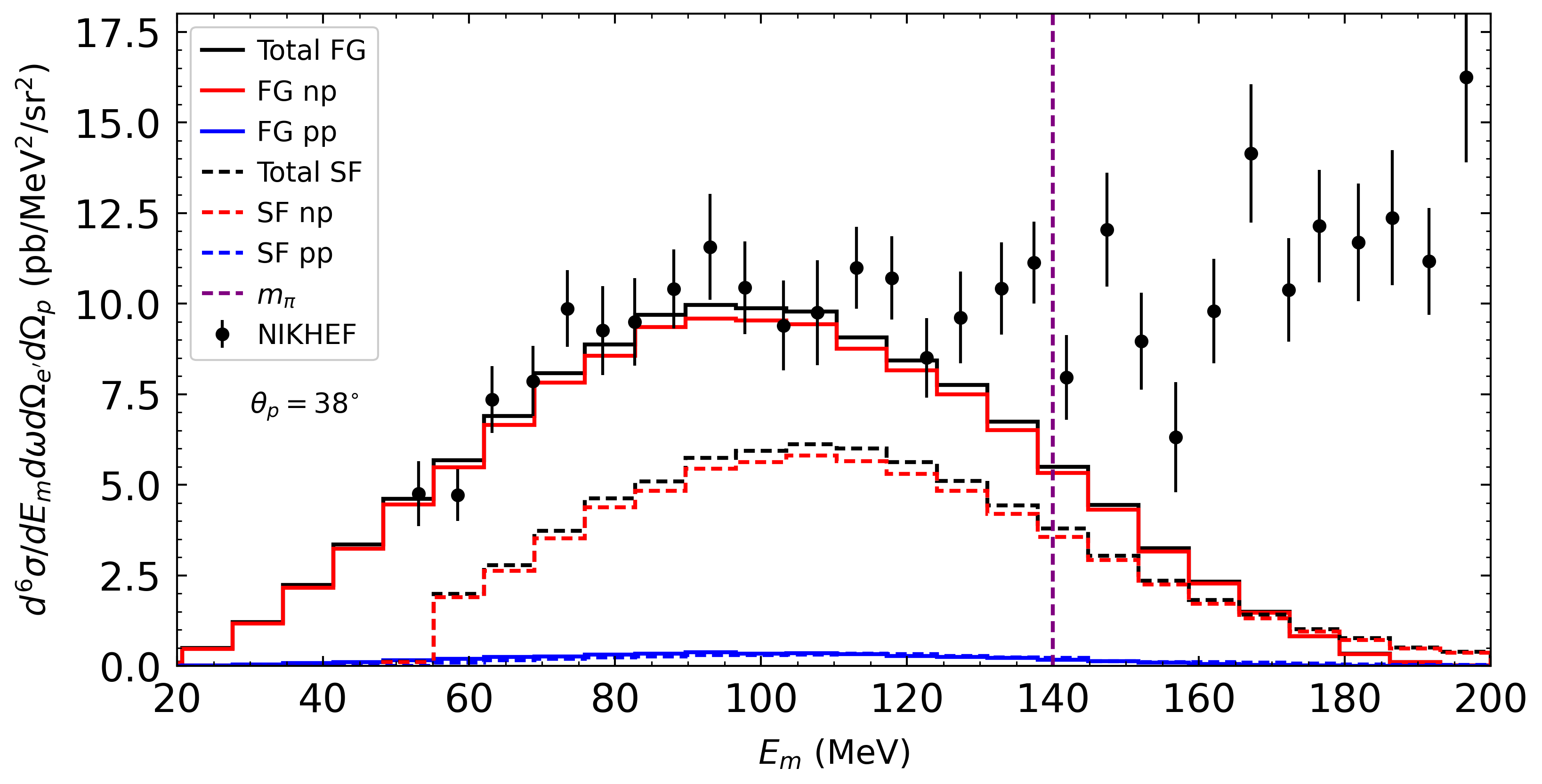}
         \label{fig:1a}
     \end{subcaption}
     \begin{subcaption}
         \centering
         \includegraphics[width=\linewidth]{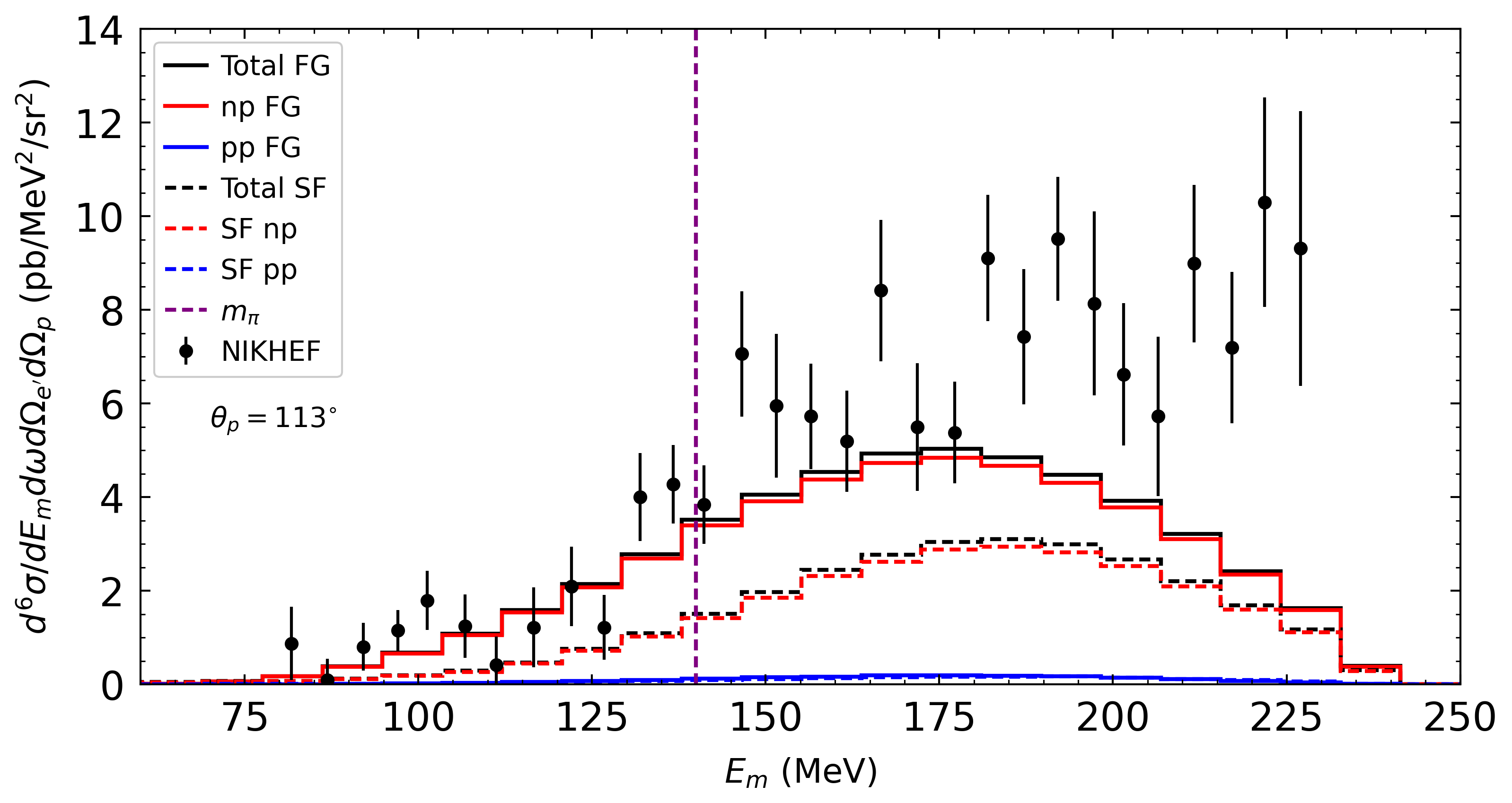}
         \label{fig:1b}
     \end{subcaption}
     \caption{Meson-Exchange contribution to the ${}^{12}{\rm C}(e,e^{\prime}p)$ cross section vs. missing energy is compared to NIKHEF data at $E_{e} = 478$ MeV, $q = 303$ MeV, $\omega = 263$ MeV and $\theta_{p} = 38^{\circ}$ (left) and $\theta_{p} = 113^{\circ}$ (right). The total contribution is split into $np$ pairs (red) and $pp$ pairs (black) with the FG (solid) and SF (dashed). Also shown as a vertical dashed line is the pion mass at $\approx140$ MeV.}
     \label{fig:eep}
\end{figure}

In this section, we apply the theoretical framework discussed above to compute a variety of semi-exclusive distributions. In order to do it efficiently, we developed a Monte Carlo generator which samples the full kinematics of the initial nuclear constituents as well as of the leptonic and hadronic particles in the final state. Event generation is based on an importance-sampling strategy guided by the two-nucleon spectral function, which ensures efficient exploration of the relevant regions of phase space and avoids spending computational effort on configurations with negligible weight~\cite{Rocco:2018mwt}. For a given incoming flux, either mono-energetic or energy-distributed, a preliminary (warm-up) run is performed to compute the total cross section—flux-averaged when appropriate—and to build the cross-section distribution required for the unweighting procedure. A second run then generates physical events using the chosen unweighting scheme. Differential distributions are then obtained by binning the (un)weighted events in the kinematic variables of interest. These results are compared with available electron-scattering data, and we also present selected distributions for neutrino scattering on the same target, for which experimental data are not currently available.

Figure~\ref{fig:eep} shows a comparison of the MEC contribution to the ${}^{12}C(e,e^{\prime}p)$ semi-exclusive cross section versus missing energy, which is defined as
\begin{equation}
    E_{m} = \omega - T_{p}\, ,
\end{equation}
with $\omega$ the energy transfer, and $T_{p}$ the kinetic energy of the detected proton. We have neglected the kinetic energy of the recoiling $A-2$ nucleus, which is a good approximation for medium-mass nuclei. The predictions are compared to NIKHEF data at $E_{e} = 478$ MeV, $q = 303$ MeV, $\omega = 263$ MeV and $\theta_{p} = 38^{\circ}$, $113^{\circ}$~\cite{Ryckebusch:1994pi}. All predictions are computed using in-plane kinematics and with $\phi_{p} = 0^{\circ}$. We follow Ref.~\cite{Belocchi:2024rfp} and use the symmetry properties of the 2p2h matrix element in Eq.~\eqref{eq:had_tensor} to compute the $(e,e^{\prime}p)$ cross section assuming the detected particle is particle 1. We sum over all isospin combinations that yield at least one proton in the final state, even if particle 1 is not itself a proton. Here we use $k_{F} = 228$ MeV, and in the case of Fermi-gas use $E_{2p2h} = 40$ MeV to compare with previous Fermi-gas calculations.

The solid lines in Fig.~\ref{fig:eep} correctly reproduce the results of Ref.~\cite{Belocchi:2024rfp} obtained within the Fermi-gas approximation, using the same $\Delta$ propagator and the same choice for the sign of the $\pi$–$\Delta$ interference. This agreement validates both our implementation and the Monte Carlo event generator developed in this work.

\begin{figure*}[htp!]
    \centering
    \includegraphics[width=0.95\textwidth]{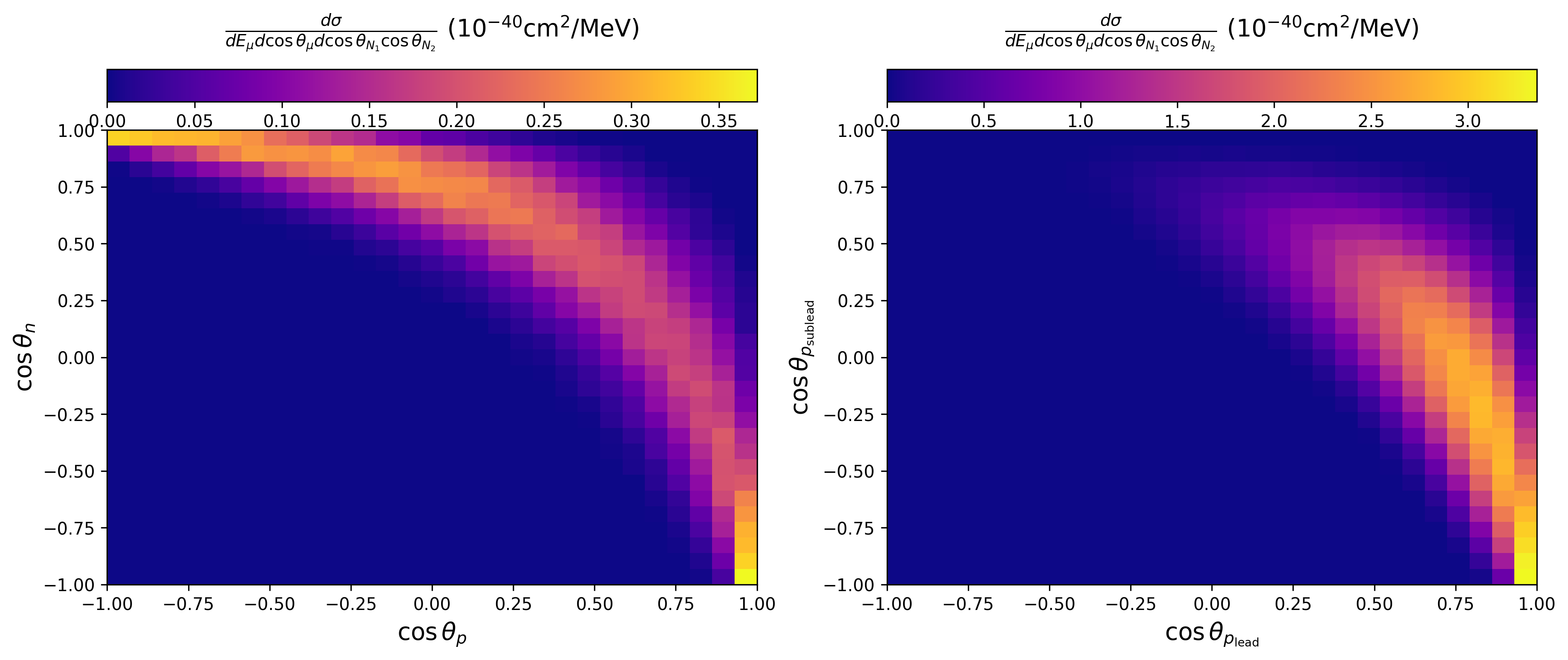}
    \caption{Fermi Gas differential cross section in the cosine of the outgoing nucleon angles for $np$ (left) and $pp$ (right) pairs}
    \label{fig:FG_fixedw_angles}
\end{figure*}
The dashed lines displays our results obtained using the two-nucleon SF. Although the agreement with the experimental data appears better for the FG calculation than for the SF result, we caution against drawing strong conclusions from this comparison. The FG model is based on an oversimplified framework with tunable parameters. Neglecting most of the nuclear correlations both in the initial and final state, it can effectively absorb the impact of several competing nuclear effects. 
As discussed in Ref.~\cite{Belocchi:2024rfp}, the level of agreement between the FG calculation and the data is strongly dependent on the specific kinematics considered. It is also important to note that the experimental data include contributions from additional reaction mechanisms that are not incorporated in the present theoretical calculations, such as quasielastic scattering with accompanied by the emission of correlated nucleon pairs. Moreover, above the pion-production threshold, also quasi-free pion production via intermediate $\Delta$ excitation
is a significant source of $(e, e^\prime p)$ strength, whose tails are likely to overlap with the meson-exchange current contribution. In this context, an underestimation of the data should not be interpreted as a deficiency of the model, but rather as a consequence of reaction mechanisms that are not included in the theoretical description.
Finally, the contribution of the $\Delta$ resonance to the two-body current is poorly constrained as discussed in Ref.~\cite{Simons:2022ltq}: different choices of model parameters, as well as alternative forms of the $\Delta$ propagator, can significantly alter the magnitude of the predicted response. A detailed assessment of the associated theoretical uncertainties—arising from the choice of $\Delta$ form factors, propagator prescriptions, $\pi$–$\Delta$ interference, and other model parameters—is deferred to future work.

The agreement observed for the other kinematic settings considered in Ref.~\cite{Belocchi:2024rfp} is similarly good, further validating the Monte Carlo event generator developed in this work.

To better understand the differences between the FG and SF initial states, and to elucidate the exclusive kinematics of MEC induced neutrino interactions, we first consider mono-energetic scattering at a fixed angle and outgoing muon energy. We choose $E_{\nu} = 950$ MeV, $\cos\theta_{\mu} = 0.85$ and $E_{\mu} = 600$ MeV. This point is near the rising edge of the $\Delta$ peak and so the total cross section should be dominated by excitations of the $\Delta$. Otherwise the kinematic point we have chosen is arbitrary and not exhaustive, but is a good representation of kinematics covered by many neutrino experiments. 

\begin{figure*}[htpb!]
    \centering
    \includegraphics[width=0.95\textwidth]{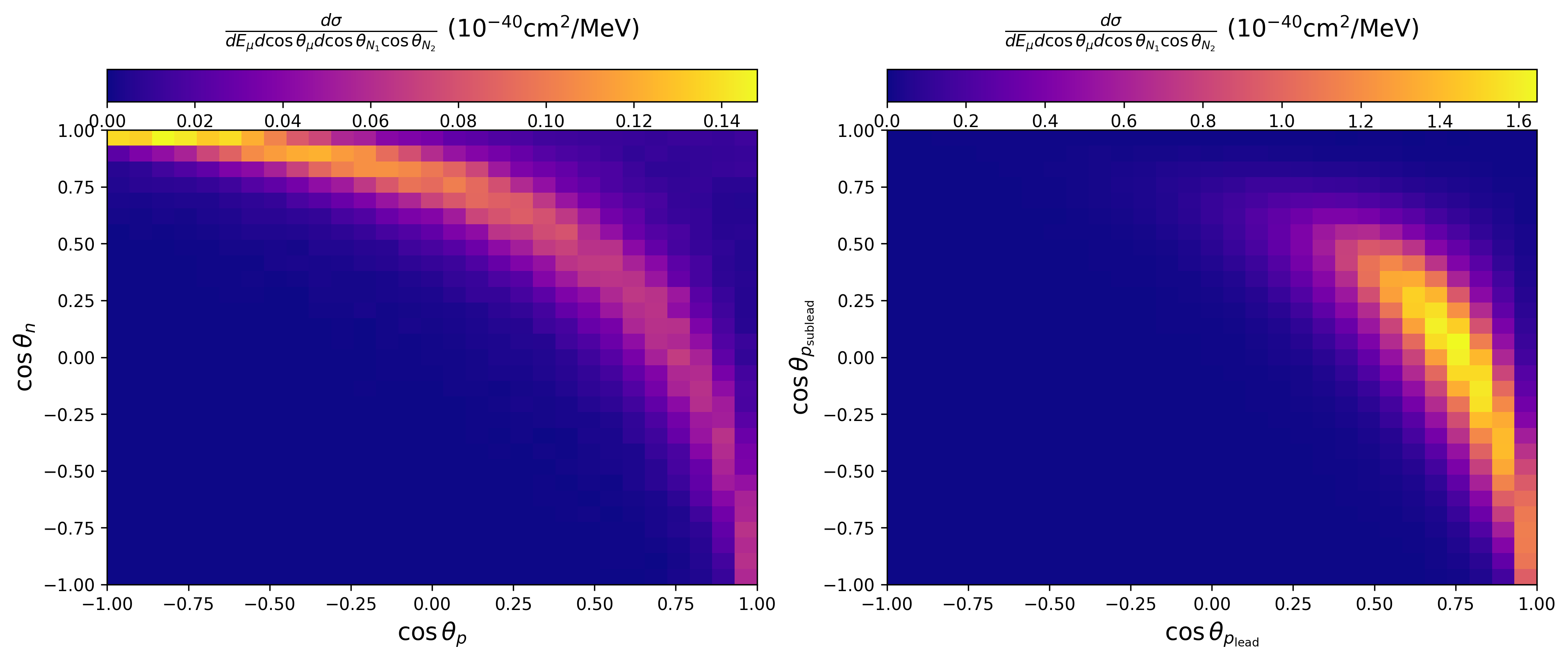}
    \caption{Spectral Function differential cross section in the cosine of the outgoing nucleon angles for $np$ (left) and $pp$ (right) pairs}
    \label{fig:SF_fixedw_angles}
\end{figure*}

Figures~\ref{fig:FG_fixedw_angles} and~\ref{fig:SF_fixedw_angles} show for the FG and SF, respectively, the differential cross sections in the cosine of the angles of the outgoing nucleons for MEC induced $\nu_{\mu}$ interactions on ${}^{12}\rm{C}$. These are further separated into outgoing $np$ and $pp$ pairs, where in the $pp$ case the leading and sub-leading protons are differentiated, breaking the symmetry being the outgoing protons. Note that for neutrinos, MEC interactions always lead to the emission of at least one proton.

In all cases it can be seen that the cross sections trace out curves which favor the emission of nucleons in relatively back-to-back configurations, reminiscent of the $\it{hammer}$ events seen at the ArgoNeuT experiment~\cite{ArgoNeuT:2014ihi}. 
The fraction of events with a back-to-back topology is marginally higher for $pp$ pairs than for $np$ pairs. In general the cross section for the FG is more broadly spread in $\cos\theta_{N_{1}},\cos\theta_{N_{2}}$ space than in SF case, as the FG is completely isotropic while the narrowness of the SF cross section reflects the correlations between $q_{rel}$ and $Q_{tot}$.

Looking at the FG case in Fig.~\ref{fig:FG_fixedw_angles}, we see that while the leading proton in both $np$ and $pp$ emission tends to be forward, there is still significant strength for backward emission in the $np$ case. In the SF distributions the direction of leading proton is less peaked, considerably so in the case of $np$ emission. The direction of the sub-leading nucleon is highly correlated with the leading nucleon, as was already discussed.

Moving from mono-energetic neutrinos with fixed kinematics to flux-folded distributions, we expect to see smaller differences between the FG and SF results, as the differences are smoothed out as one integrates over the broad experimental neutrino flux. The results presented below focus mostly on distributions of the leading proton, reflecting the experimental practice of reporting cross sections for the $Np0\pi$ topology, in which, for events with $N>1$, only the kinematics of the leading proton are used in the analysis. We focus here on the T2K experiment as it's beam peaks in a region where MEC interactions still make up a substantial portion of the total cross section.

\begin{figure}
     \centering
     \includegraphics[width=\linewidth]{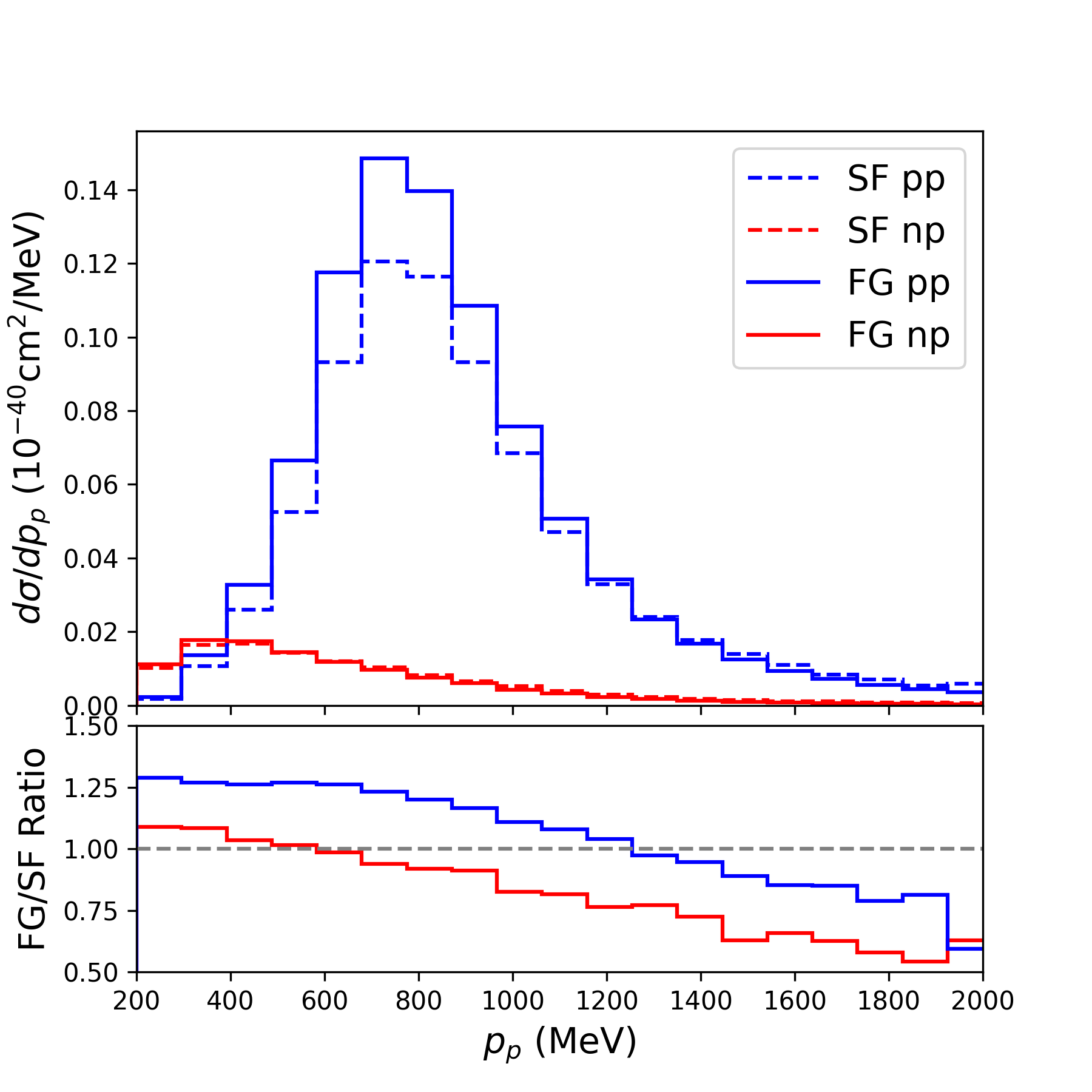}
     
     \caption{Flux-folded differential cross section for $\nu_{\mu}$ on ${}^{12}\rm{C}$ in $p_{p}$ of the leading proton at T2K for both outgoing $pp$ pairs (blue) and $np$ pairs (red). The FG results correspond to the solid lines, while those with the SF are the dashed lines. The bottom plot shows the ratio of the FG to SF prediction for both $pp$ and $np$ pairs.}
     \label{fig:T2K_pplead}
\end{figure}

The T2K experiment employs a relatively narrow neutrino beam with a mean energy peaked around 600 MeV. The hard scattering processes most relevant for T2K are quasi-elastic scattering and meson-exchange currents. Figure~\ref{fig:T2K_pplead} shows the momentum distribution of the leading proton, separated into contributions from $pp$ and $np$ pairs. For $pp$ pairs, both the FG and SF calculations peak at momenta around 800 MeV, whereas protons originating from $np$ pairs peak at 400 MeV. 

In both $pp$ and $np$ channels, the SF calculation yields an enhanced population of high-momentum protons, reflecting the high-momentum tail of the input spectral function. Notably, the ratio between the FG and SF results differ for $pp$ and $np$ pairs: in the SF case, the number of $pp$ pairs at low leading proton momentum is reduced by up to 20\% relative to the FG prediction.

This difference has important implications for measurements of the $Np$ final-state topology, as $pp$ pairs constitute the dominant contribution to the meson-exchange current cross section. As discussed below, a similar pattern is observed for the other kinematic distributions.

\begin{figure}
     \centering
     \includegraphics[width=\linewidth]{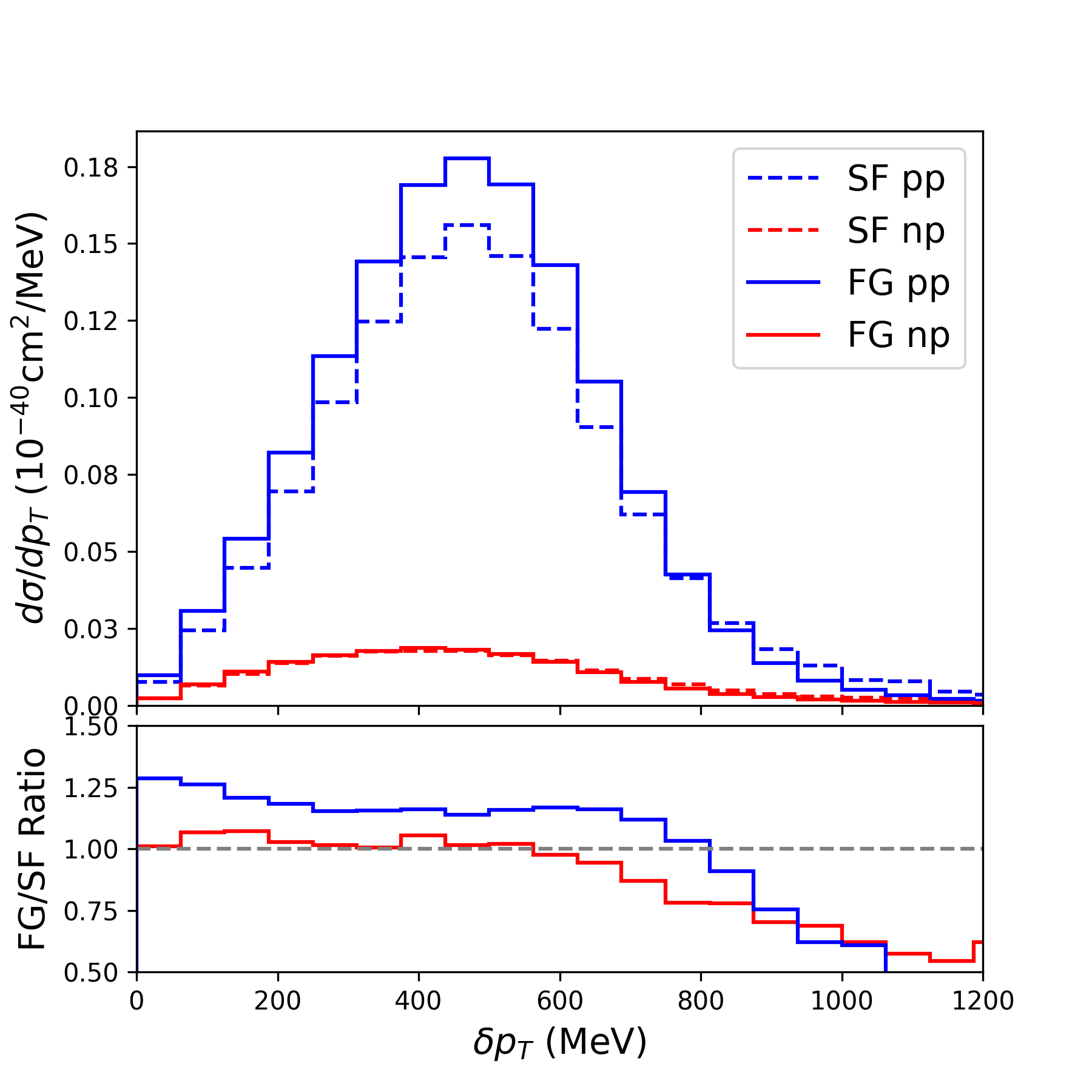}
     
     \caption{Flux-folded differential cross section for $\nu_{\mu}$ on ${}^{12}\rm{C}$ in $\delta p_{T}$ including only the leading proton at T2K for both outgoing $pp$ pairs (blue) and $np$ pairs (red). The FG results correspond to the solid lines, while those with the SF are the dashed lines. The bottom plot shows the ratio of the FG to SF prediction for both $pp$ and $np$ pairs.}
     \label{fig:T2K_pT}
\end{figure}

Similar results can be drawn from Fig.~\ref{fig:T2K_pT} showing the $\delta p_{T}$ of the leading proton-muon system. Here the distribution of $\delta p_{T}$ for $np$ pairs in the SF and FG results are very similar except for large values of $\delta p_{T}$, while the number of $pp$ pairs in the peak for the FG case is 20\% larger than the SF case.

\begin{figure}
     \centering
     \includegraphics[width=\linewidth]{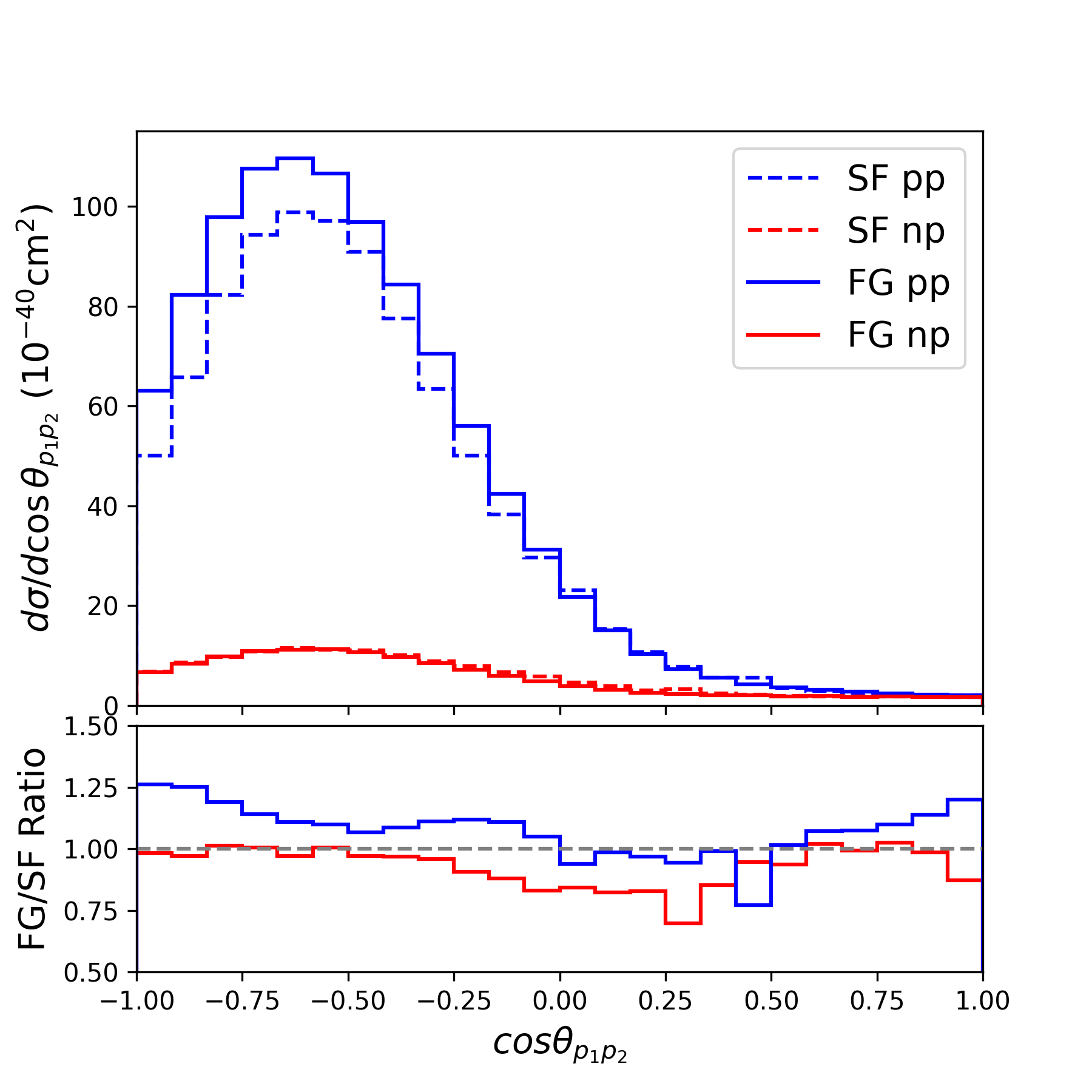}
     
     \caption{Flux-folded differential cross section for $\nu_{\mu}$ on ${}^{12}\rm{C}$ in the angle between outgoing nucleons at T2K for both outgoing $pp$ pairs (blue) and $np$ pairs (red). The FG results correspond to the solid lines, while those with the SF are the dashed lines. The bottom plot shows the ratio of the FG to SF prediction for both $pp$ and $np$ pairs.}
     \label{fig:T2K_cosp1p2}
\end{figure}

The final distribution of interest is the cosine of the opening angle between the outgoing nucleons $\cos\theta_{p_{1}p_{2}}$ displayed in Fig.~\ref{fig:T2K_cosp1p2}. Though T2K has not measured this distribution for multi-nucleon events, experiments like MicroBooNE, SBND, and Icarus have or are planning to measure this quantity in two-proton knockout events~\cite{MicroBooNE:2022emb}. For these $pp$ events, the FG predicts more events at backwards angles while for $np$ events the distribution of angles between outgoing nucleons is very similar for the FG and SF. This constitutes the first microscopic prediction of this observable and is made possible by our fully exclusive implementation of meson-exchange currents.

\section{Conclusions}


In this work, we make several improvements to the SF formalism to accurately model two-nucleon knockout reactions induced by MEC. We employ a two-nucleon SF obtained from VMC calculations, retaining information on the angular correlations between the nucleons. Moreover, we updated the current operators used to describe the $\Delta$ contribution by adopting updated axial and vector form factors and an improved parameterization of the $\Delta$ propagator.

We compare our predictions with semi-exclusive electron--carbon scattering data and validate our framework by recovering the Fermi-gas results presented in Ref.~\cite{Belocchi:2024rfp}. To obtain these results, we developed a Monte Carlo event generator that provides the complete kinematics of both outgoing nucleons and the outgoing lepton on an event-by-event basis. This generator is ideally suited for producing multidimensional distributions of experimental interest. Moreover, after incorporating intranuclear cascade effects, the generated events can be directly interfaced with experimental simulation frameworks to predict the full detector response to meson-exchange current processes.

We consider the electroweak case and present results for mono-energetic and flux-folded neutrino cross sections on a carbon target. We observe that the cross sections favor the emission of nucleons in predominantly back-to-back configurations. In general, the Fermi-gas predictions plotted as a function of the cosine of the outgoing nucleons angle are more broadly distributed than those obtained using the SF, reflecting the isotropic nature of the Fermi-gas model and the correlations between $q_{\mathrm{rel}}$ and $Q_{\mathrm{tot}}$ present in the spectral-function approach.

We also observe differences in the leading-proton momentum distributions for both $np$ and $pp$ emission channels when comparing the Fermi-gas and spectral-function calculations, indicating that nucleon--nucleon correlations play an important role in reproducing experimentally measured distributions in neutrino scattering. Focusing on flux-folded cross sections using the T2K neutrino flux, we find that, in both $pp$ and $np$ channels, the spectral-function calculation yields an enhanced population of high-momentum protons, reflecting the high-momentum tail of the underlying nuclear spectral function. Notably, the ratio between the FG and SF results differs for $pp$ and $np$ pairs: in the SF case, the number of $pp$ pairs at low leading-proton momentum is reduced by up to 20\% relative to the Fermi-gas prediction. This difference has important implications for measurements of the $Np$ final-state topology, as $pp$ pairs constitute the dominant contribution to the meson-exchange current cross section. A similar pattern is observed for the $\delta p_T$ distribution and for the cosine of the opening angle between the outgoing nucleons.

The results presented in this work represent a first step toward a systematic comparison with electron- and neutrino-scattering data in kinematic regimes where two-nucleon emission is relevant. Future developments will include a detailed assessment of theoretical uncertainties associated with form factors and coupling constants, an investigation of final-state interactions, and extensions of the nuclear calculations to heavier targets. Our findings already indicate that the structure of the $\Delta$ current and nucleon--nucleon correlations, particularly angular correlations, have a measurable impact on observable quantities.

\begin{acknowledgments}

We would like to thank Valerio Belocchi, Juan Nieves and Joanna Sobczyk for providing useful insights about the current operators, as well as Joshua Isaacson for insightful discussions on unweighting and event generation. We would like to thank Alessandro Lovato for reading the manuscripts. This manuscript has been authored by Fermi Forward Discovery Group, LLC under Contract No. 89243024CSC000002 with the U.S.\ Department of Energy, Office of Science, Office of High Energy Physics.
The present research is supported by the SciDAC-5 NeuCol program (N.~R. and N.~S.), by the Neutrino Theory Network (A.~N.), and by the Nuclear Theory for New Physics Topical collaboration (N.~S.).

This work was supported in part by IFIC and Universitat de Valencia (N.~R.).

\end{acknowledgments}

\appendix

\section{Numerical Integration}

In this Section we outline how the numerical integration over the initial- and final-state nucleon phase space is performed.  
The integration over the final-state nucleon momenta is carried out most conveniently in the center-of-momentum (CM) frame of the outgoing pair.  
In that frame one samples the solid angle of one final-state nucleon at random, while its energy and momentum are fixed by energy--momentum conservation.  
The corresponding four-momenta are then boosted back to the laboratory frame.

We start from the integral
\begin{align}
    \mathcal{I}
    &= \int d^{3}p_{1}\, d^{3}p_{2}\;
    \frac{1}{E_{p_1} E_{p_2}}\,
    \delta\!\left(E_{\mathrm{tot}} - E_{p_1} - E_{p_2}\right)\nonumber\\
    &\times \delta^{(3)}\!\left(\mathbf{p}_{\mathrm{tot}} - \mathbf{p}_{1} - \mathbf{p}_{2}\right)
    \left|\mathcal{M}\right|^{2}\,,
\end{align}
where \(\mathbf{p}_{1}\) and \(\mathbf{p}_{2}\) are the final nucleon momenta with corresponding energies
\(E_{p_1}\) and \(E_{p_2}\), 
\begin{equation}
    E_{\mathrm{tot}} = \omega - E + 2m_{N}\,,
    \qquad
    \mathbf{p}_{\mathrm{tot}} = \mathbf{h}_{1} + \mathbf{h}_{2} + \mathbf{q}\,,
\end{equation}
and $|\mathcal{M}|^2$ denotes the full matrix element squared.  

We now boost into the center-of-momentum frame of the final-state nucleons, where
\begin{equation}
    \mathbf{p}_{\mathrm{tot}}^{\prime\prime}
    = \mathbf{p}_{1}^{\prime\prime} + \mathbf{p}_{2}^{\prime\prime}
    = \mathbf{0}\,.
\end{equation}
Quantities in this frame are denoted by a double prime.  
The total energy in the CM frame is fixed by Lorentz invariance,
\begin{equation}
    E_{\mathrm{tot}}^{\prime\prime}
    = \sqrt{E_{\mathrm{tot}}^{2} - \mathbf{p}_{\mathrm{tot}}^{2}}\,.
\end{equation}
Neglecting the neutron–proton mass difference, the two nucleons in the CM frame move back-to-back with equal energy,
\begin{equation}
    E_{1}^{\prime\prime} = E_{2}^{\prime\prime}
    = \frac{E_{\mathrm{tot}}^{\prime\prime}}{2}\,.
\end{equation}

Using the Lorentz invariance of the phase-space measure,
\begin{equation}
    \frac{d^{3}p_{i}}{E_{i}}
    = \frac{d^{3}p_{i}^{\prime\prime}}{E_{i}^{\prime\prime}}\,,
\end{equation}
we can rewrite the integral in the CM frame and use the three-momentum delta function to perform the integration over \(\mathbf{p}_{2}^{\prime\prime}\).  
The integral becomes
\begin{align}
    \mathcal{I}
    &= \int d^{3}p_{1}^{\prime\prime}\;
    \frac{1}{\left(E_{1}^{\prime\prime}\right)^{2}}\,
    \delta\!\left(2E_{1}^{\prime\prime} - E_{\mathrm{tot}}^{\prime\prime}\right)\nonumber\\
    &\times\Theta\!\left(E_{\mathrm{tot}}^{2} - \mathbf{p}_{\mathrm{tot}}^{2} - 4m_{N}^{2}\right)
    \left|\mathcal{M}\right|^{2}\,,
\end{align}
where the step function \(\Theta\) enforces the kinematic threshold for production of a two-nucleon final state.

Switching to spherical coordinates in the CM frame, we write
\begin{equation}
    d^{3}p_{1}^{\prime\prime}
    = {p_{1}^{\prime\prime}}^{2}\, dp_{1}^{\prime\prime}\, d\Omega_{1}^{\prime\prime}\,,
\end{equation}
and use the relation $p_{1}^{\prime\prime} dp_{1}^{\prime\prime}= E_{1}^{\prime\prime} dE_{1}^{\prime\prime}\,.$
The integration over \(E_{1}^{\prime\prime}\) is then carried out using the delta function, which fixes
$ E_{1}^{\prime\prime}= {E_{\mathrm{tot}}^{\prime\prime}}/{2}\,$
and yields an additional factor of \(1/2\).  
The final expression for the phase-space integral is
\begin{equation}
    \mathcal{I}
    = \frac{p_{1}^{\prime\prime}}{2E_{1}^{\prime\prime}}
    \int d\Omega_{1}^{\prime\prime}\,
    \Theta\!\left(E_{\mathrm{tot}}^{2} - \mathbf{p}_{\mathrm{tot}}^{2} - 4m_{N}^{2}\right)
    \left|\mathcal{M}\right|^{2}\,,
\end{equation}
where \(p_{1}^{\prime\prime}\) and \(E_{1}^{\prime\prime}\) are evaluated at the on-shell CM value determined above.

The matrix element squared, \(|\mathcal{M}|^{2}\), is computed in the laboratory frame.  
Therefore, after sampling the final-state nucleon momenta in the CM frame, they are boosted back to the laboratory frame following the prescription of Amaro \emph{et al.}~\cite{Martinez-Consentino:2023hcx}.


\bibliographystyle{apsrev}
\bibliography{biblio}

\end{document}